\documentclass{PoS}

\newcommand{\be}{\begin{equation}}
\newcommand{\ee}{\end{equation}}

\newcommand{\bi}{\begin{itemize}}
\newcommand{\ei}{\end{itemize}}

\newcommand{\bea}{\begin{eqnarray*}}
\newcommand{\eea}{\end{eqnarray*}}

\title{Spectral Flow and Index Theorem for Staggered Fermions}

\ShortTitle{Spectral Flow and Index Theorem for Staggered Fermions}

\author{\speaker{E. Follana} \\
        Universidad de Zaragoza\\
        E-mail: \email{efollana@unizar.es}}

\author{V. Azcoiti\\
        Universidad de Zaragoza\\
        E-mail: \email{azcoiti@azcoiti.unizar.es}}

\author{G. Di Carlo\\
        INFN, Laboratori Nazionali del Gran Sasso\\
        E-mail: \email{giuseppe.dicarlo@lngs.infn.it}}

\author{A. Vaquero \footnote{Present address: The Cyprus Institute.}\\
        Universidad de Zaragoza\\
        E-mail: \email{alexv@unizar.es}}

\abstract{We investigate numerically the spectral flow introduced by
  Adams for the staggered Dirac operator on realistic gauge
  configurations. We study both the unimproved and the HISQ Dirac
  operators. We compare the spectral flow index with the index
  obtained by identifying low-lying modes of large chirality.}

\FullConference{XXIX International Symposium on Lattice Field Theory \\
		 July 10-16 2011\\
		 Squaw Valley, Lake Tahoe, California}

\begin{document}

\section{Introduction}

In \cite{Adams1}, Adams introduced a new definition of topological
charge for lattice gauge fields based on the spectral flow of a
hermitian operator related to the staggered Dirac operator. Some
numerical results were obtained there for synthetic configurations in
the $2D$ U(1) model.

Here we present preliminary\footnote{More complete results will be
  presented in \cite{futuro}.} numerical results in realistic, $4D$
pure gauge $SU(3)$ configurations, confirming the good properties of
Adams' definition, and the agreement of the index calculated with the
new definition and by counting the number of low-lying modes of high
chirality.

\section{Definition of the topological charge}

The hermitian operator introduced in \cite{Adams1} is defined by
\be
H_{st}(m) = iD_{st} - m \Gamma_5
\ee
where $D$ is the massless staggered Dirac operator and $\Gamma_5$ is
the taste-singlet staggered $\gamma_5$ \cite{Golterman}. This operator
is hermitian, and we can study its spectral flow, $\lambda(m)$. The
would-be zero modes of $D_{st}$ are now identified with the eigenmodes
for which the corresponding eigenvalue flow $\lambda(m)$ crosses zero
at low values of $m$, and the chirality of any such mode equals (with
our conventions) the sign of the slope of the crossing \cite{Adams1}.

For the most part we work with the highly improved Dirac operator
(HISQ) \cite{hisq}, although for comparison we will also show some
results corresponding to the unimproved (1-link) Dirac operator.

To compare with previous work, we also calculate the low-lying modes
of the HISQ Dirac operator at $m = 0$, and identify the would-be zero
modes with the high taste-singlet chirality ones \cite{top1, top2}.

\section{Results}

For our numerical calculations we use configurations from an ensemble
of tree-level Symanzik and tadpole improved quenched QCD with a
lattice spacing of approximately $0.077$ fm \cite{top1}.

The operator $H_{st}(m)$ is hermitian, and its low-lying eigenmodes
are easily calculated numerically with standard methods. Its spectrum
has the exact symmetry $\lambda(m) \leftrightarrow - \lambda(-m)$,
therefore we only show results for $m < 0$. An equal number of
crossings, with identical slope, will be present for $m > 0$.

\begin{figure}[h]
\begin{center}
\[
\begin{array}{cc}
\includegraphics[scale = .6]{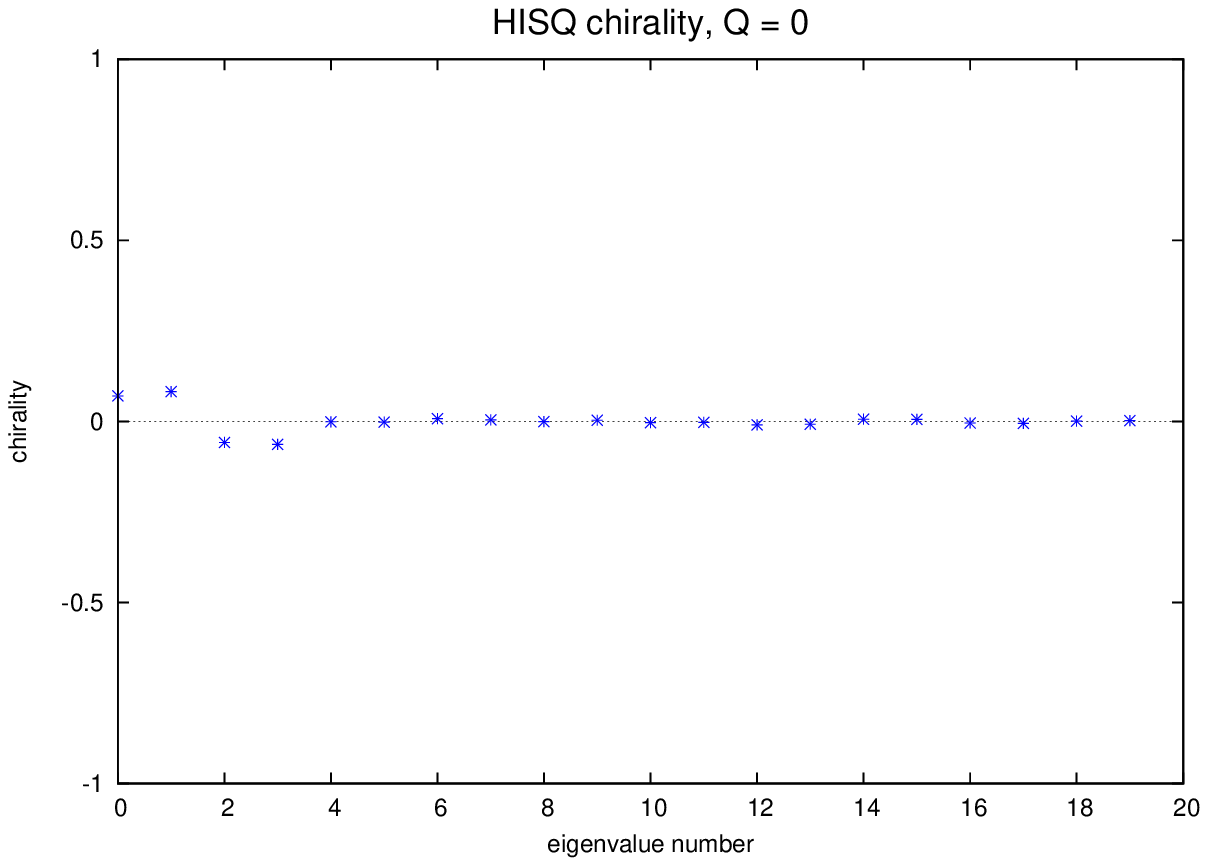} &
\includegraphics[scale = .6]{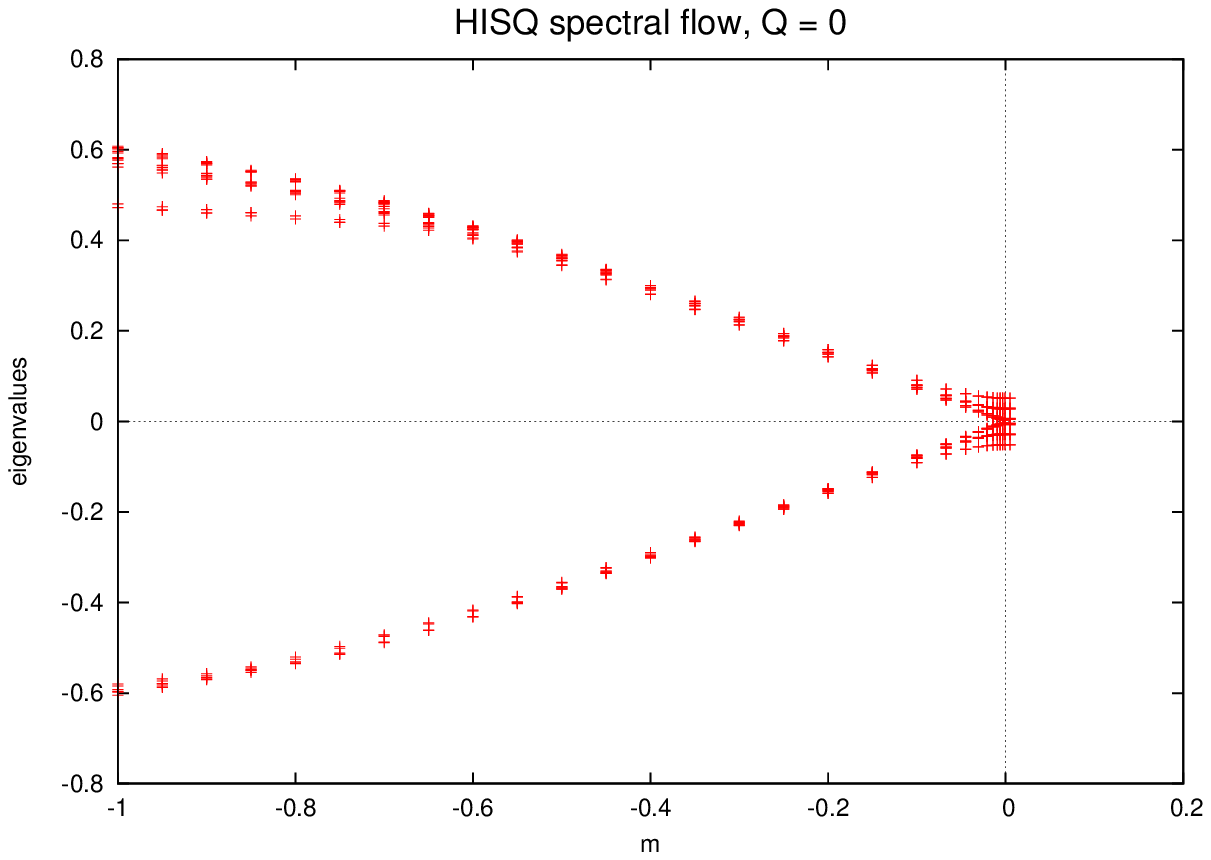} \\
\includegraphics[scale = .6]{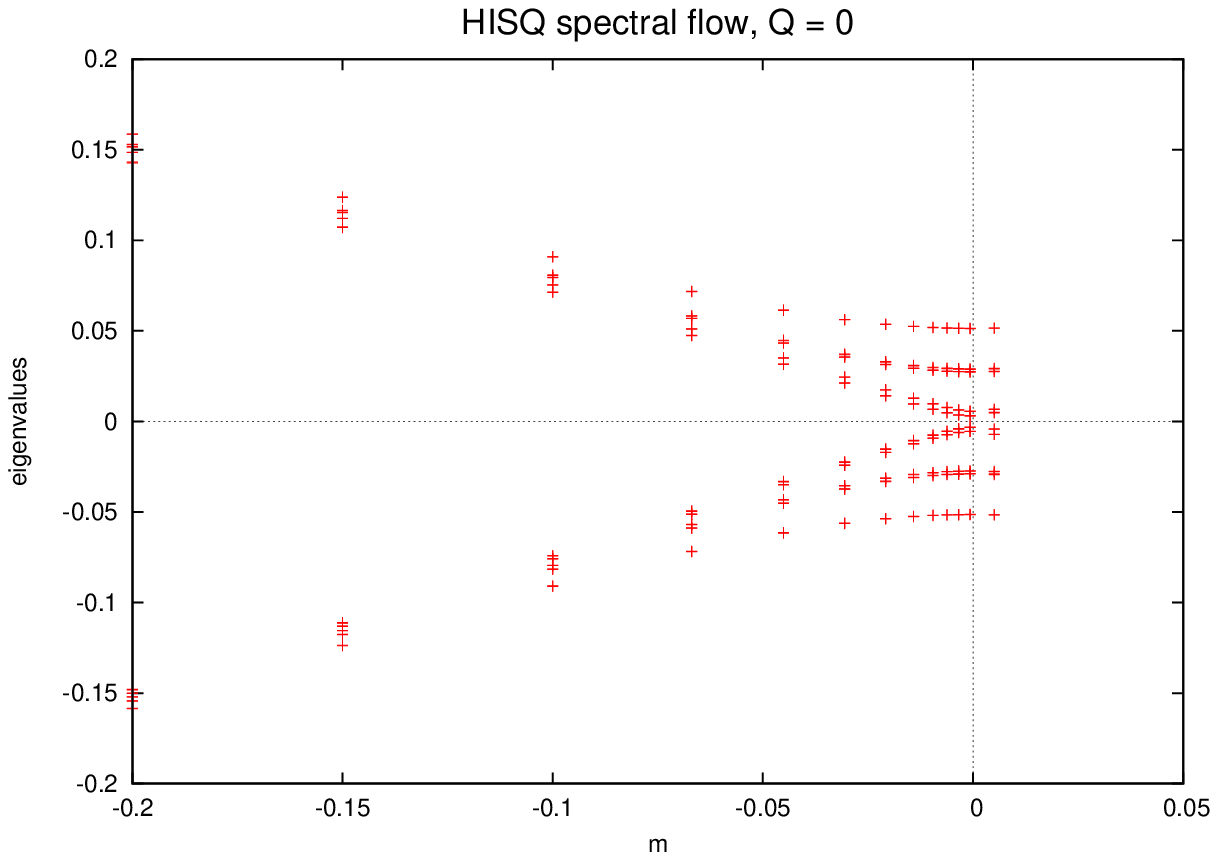} &
\includegraphics[scale = .6]{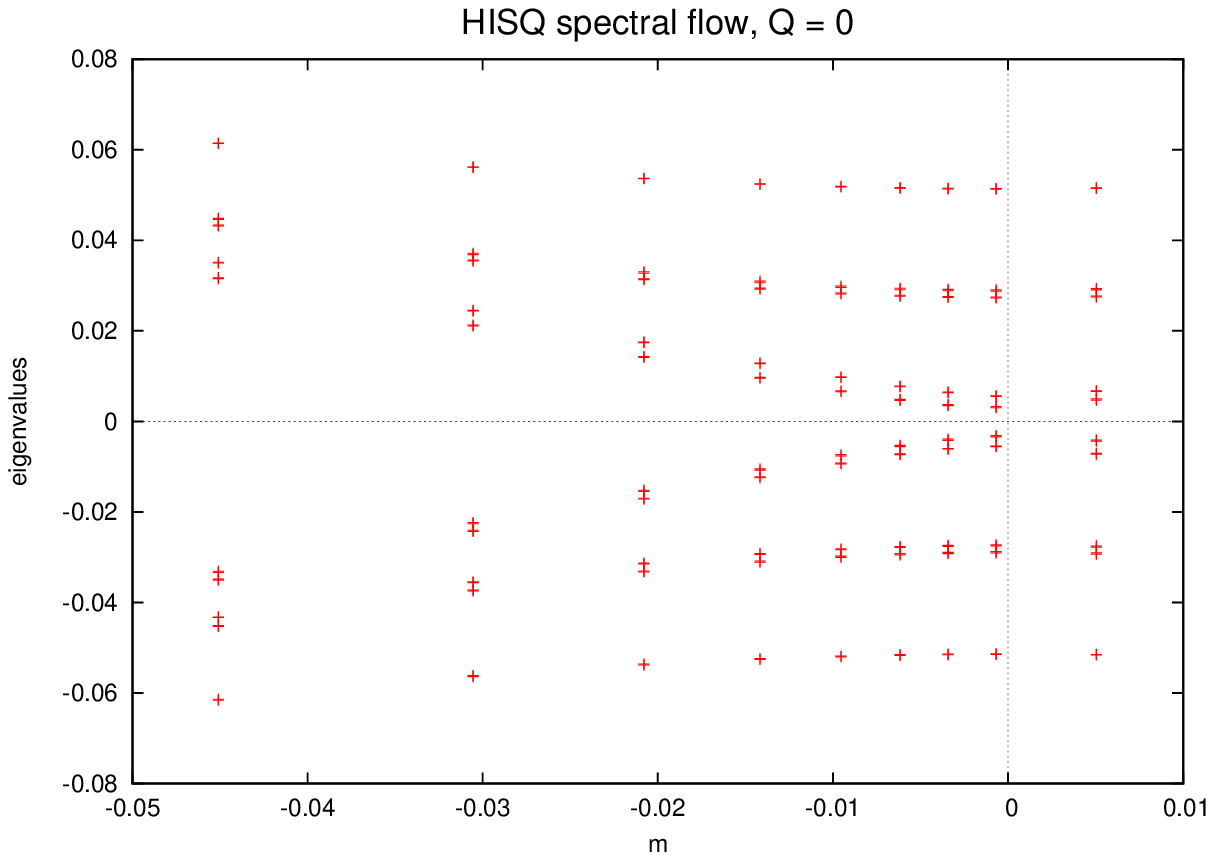} 
\end{array}
\]
\end{center}
\caption {Top left figure: taste-singlet chirality for the low-lying
  modes of the HISQ Dirac operator (only half of the modes are shown,
  as the other half is exactly degenerate, due to an exact symmetry of
  the Dirac action.) Top right and bottom figures: spectral flow for
  the low-lying modes of the corresponding hermitian operator
  $H_{st}(m)$, for various ranges of $m$ (we only show the range
  $m<0$, due to the exact symmetry $\lambda(m) \leftrightarrow
  -\lambda(-m)$). This is for a gauge configuration with $Q = 0$.}
\label{fig1}
\end{figure}

\begin{figure}[h]
\begin{center}
\[
\begin{array}{cc}
\includegraphics[scale = .6]{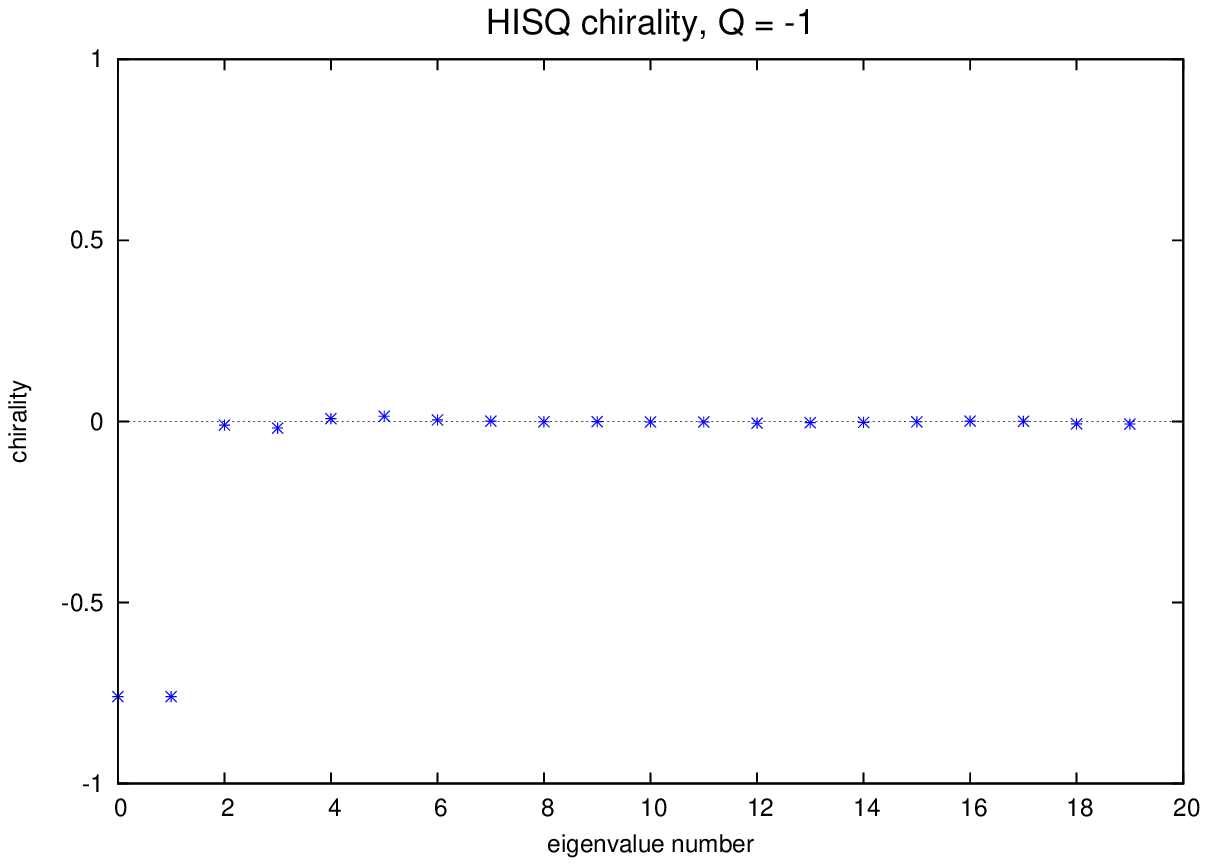} &
\includegraphics[scale = .6]{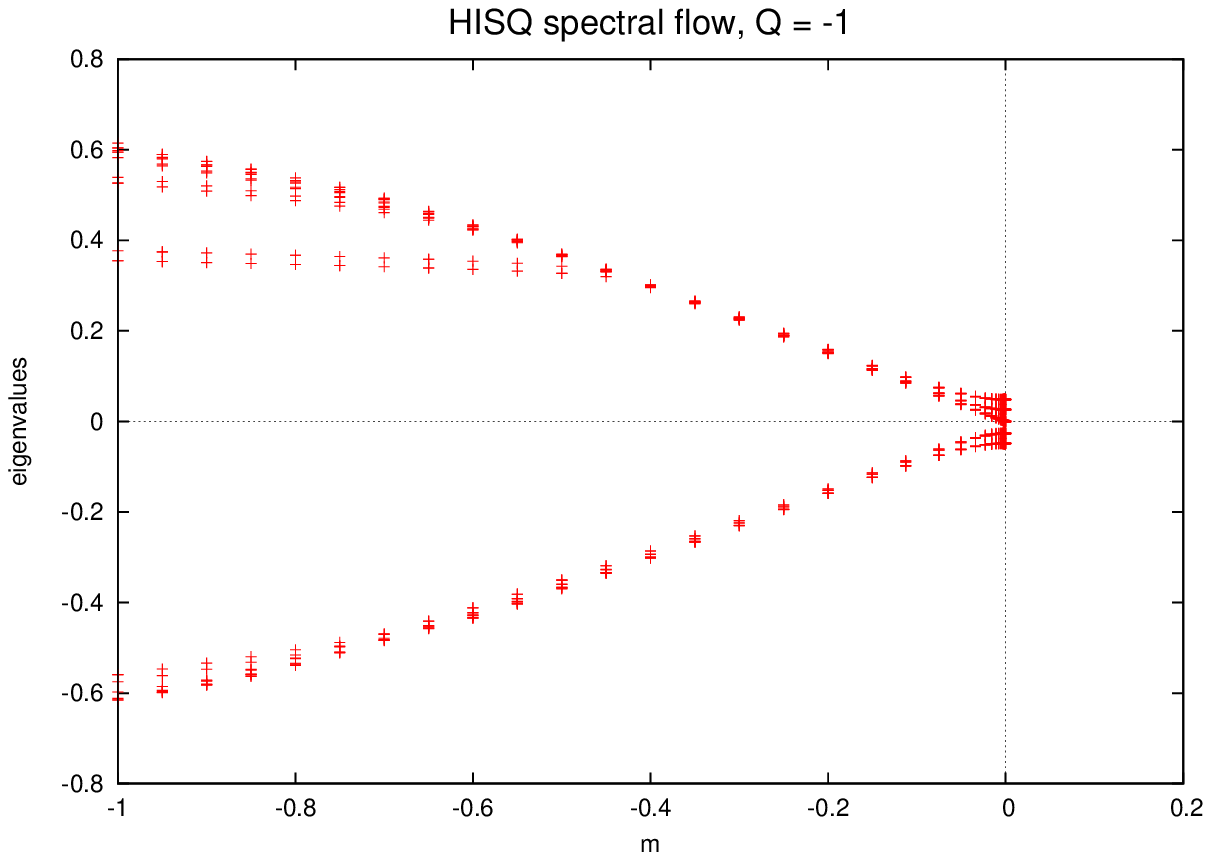} \\
\includegraphics[scale = .6]{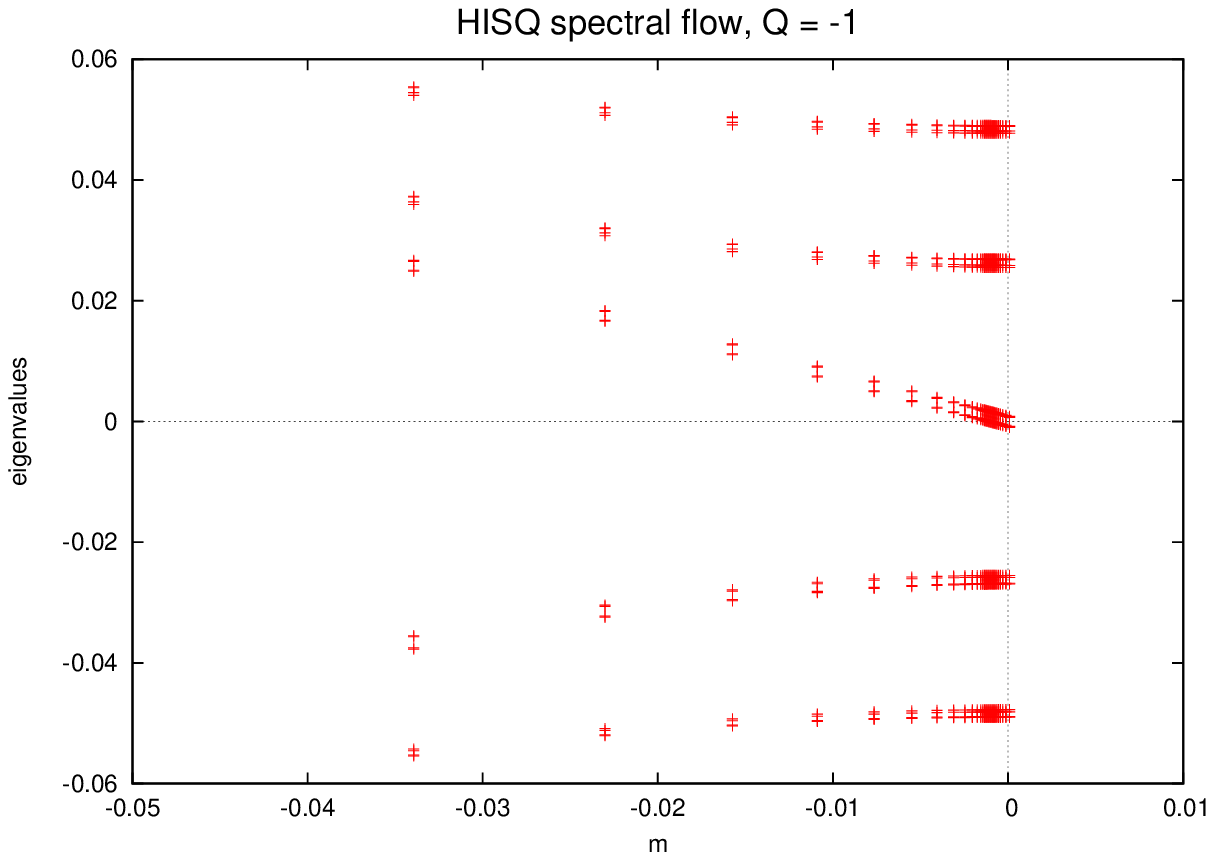} &
\includegraphics[scale = .6]{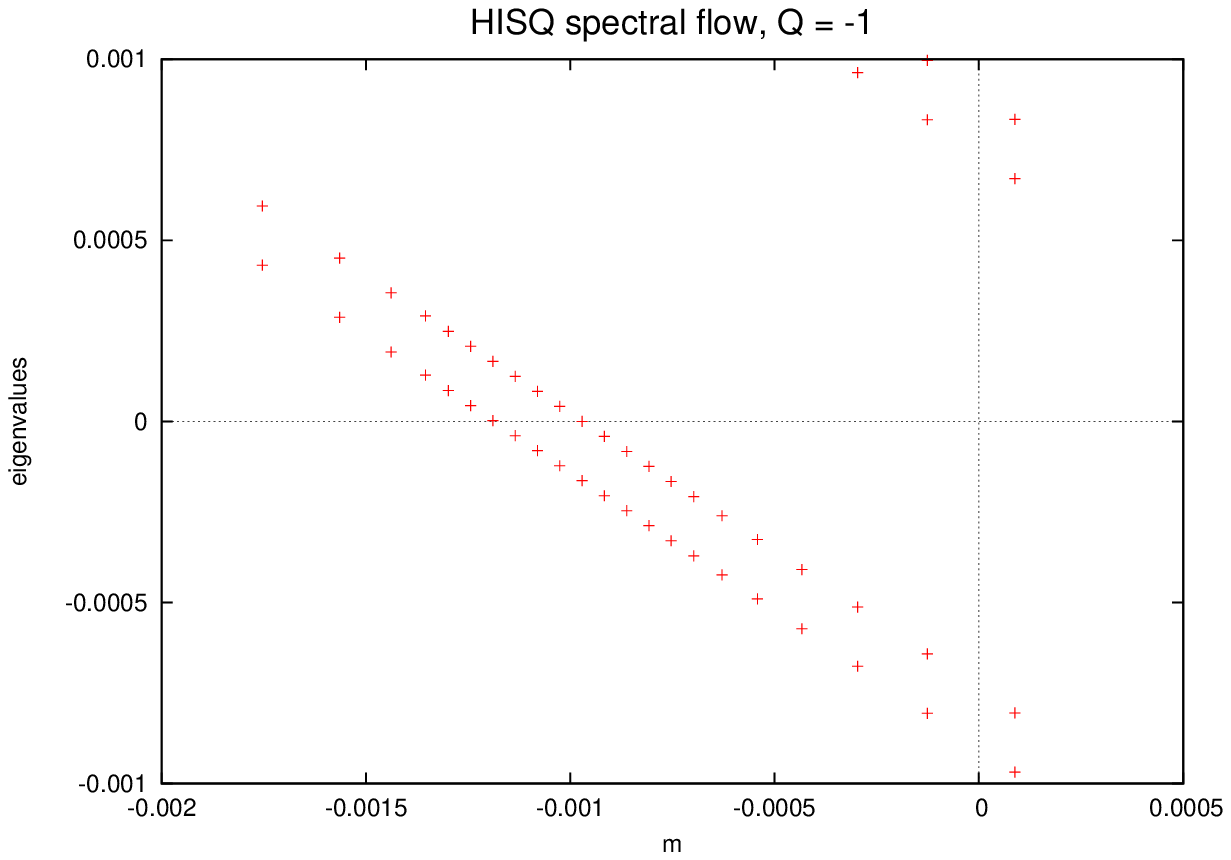} 
\end{array}
\]
\end{center}
\caption {Top left figure: taste-singlet chirality for the low-lying
  modes of the HISQ Dirac operator. Top right and bottom figures:
  spectral flow for the low-lying modes of the corresponding hermitian
  operator $H_{st}(m)$, for various ranges of $m$. This is for a gauge
  configuration with $Q = -1$.}
\label{fig2}
\end{figure}

\begin{figure}[h]
\begin{center}
\[
\begin{array}{ccc}
\includegraphics[scale = .6]{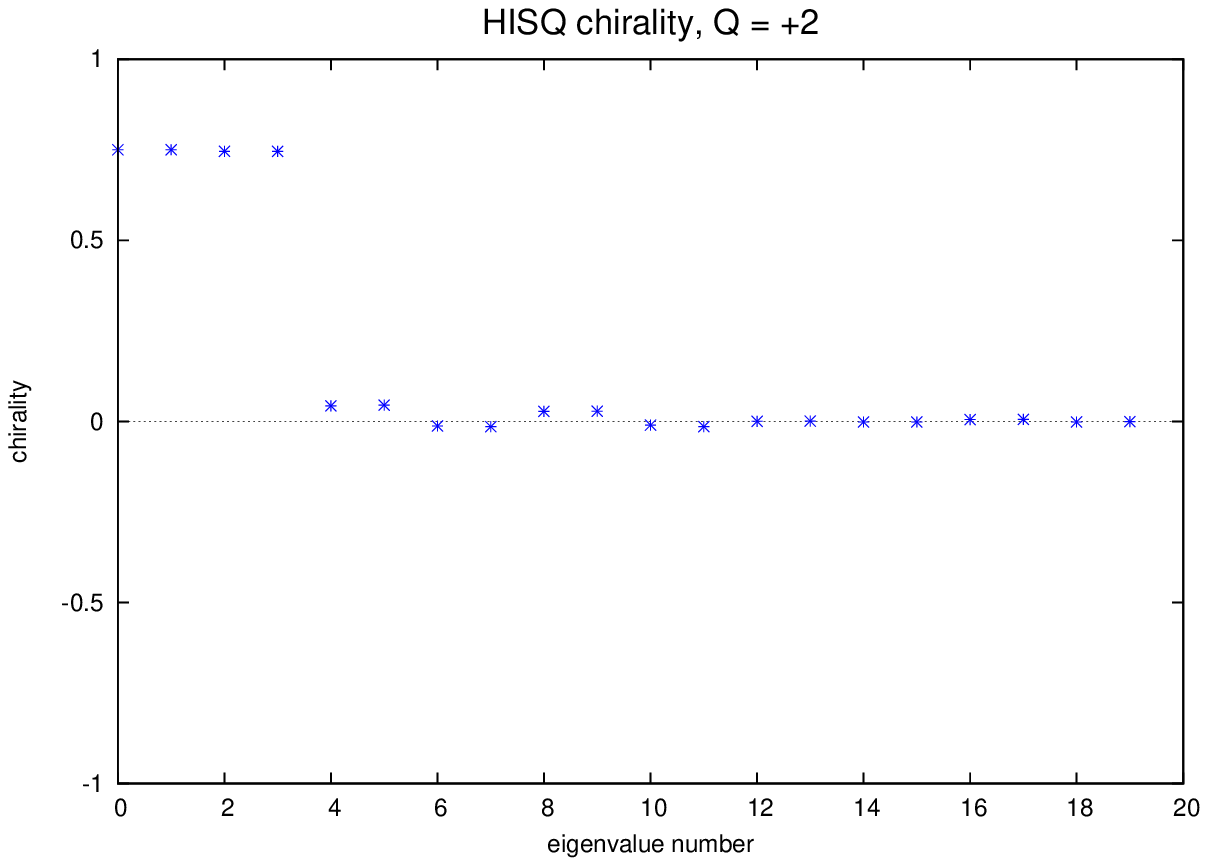} &&
\includegraphics[scale = .6]{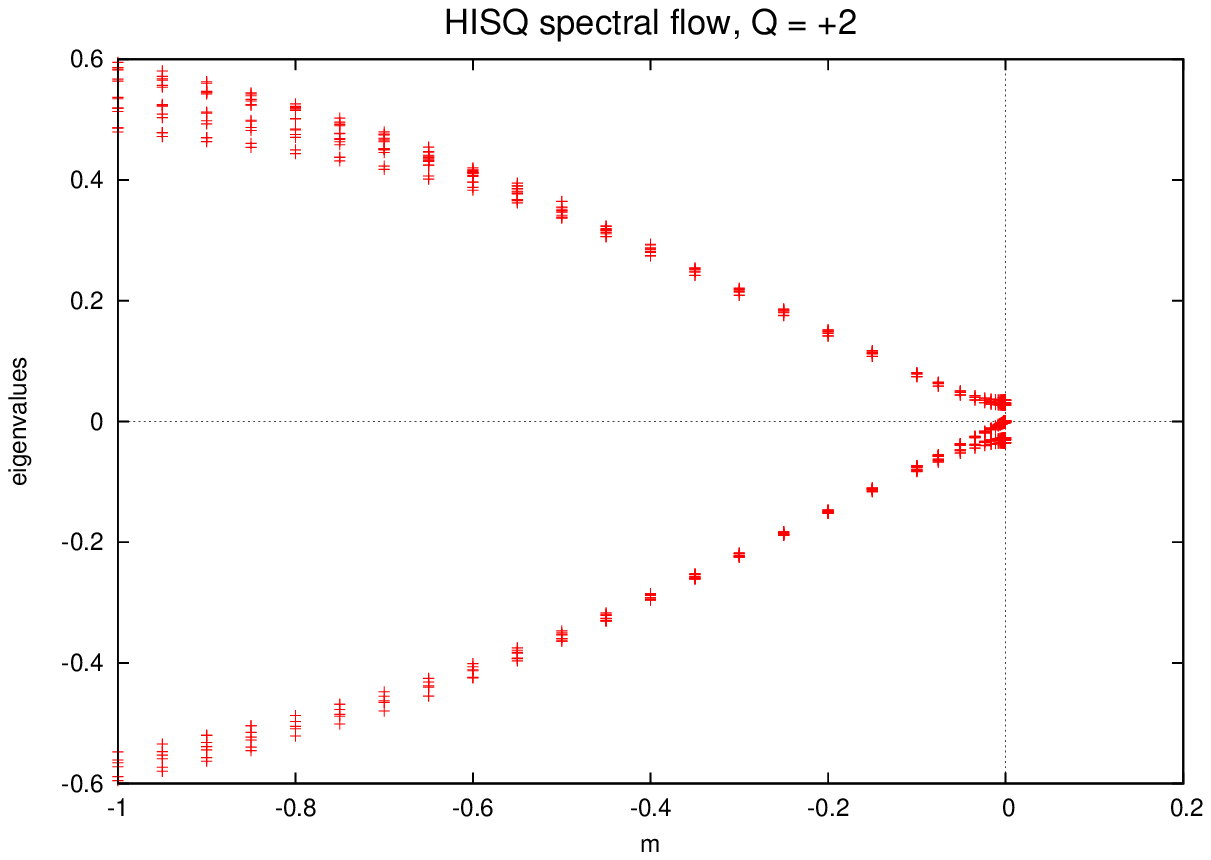} \\ 
\includegraphics[scale = .6]{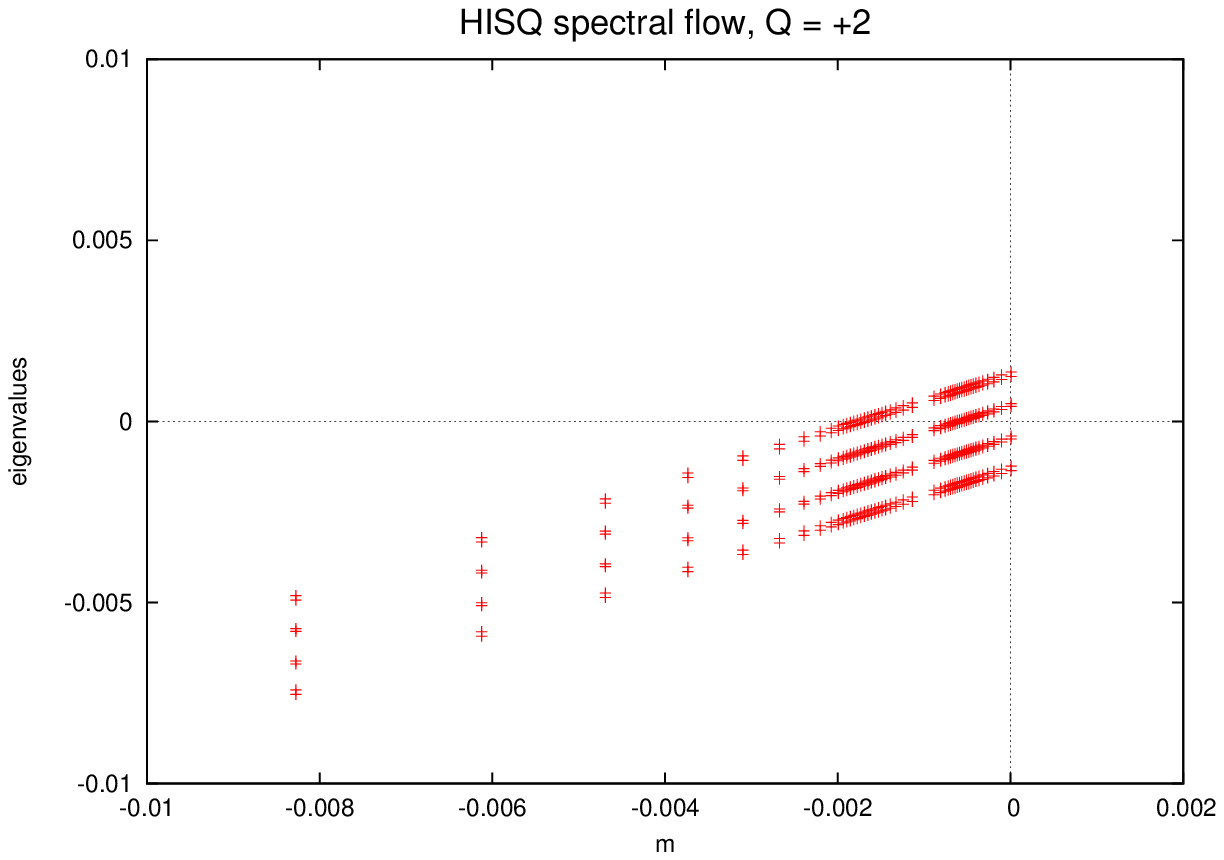} &&
\includegraphics[scale = .6]{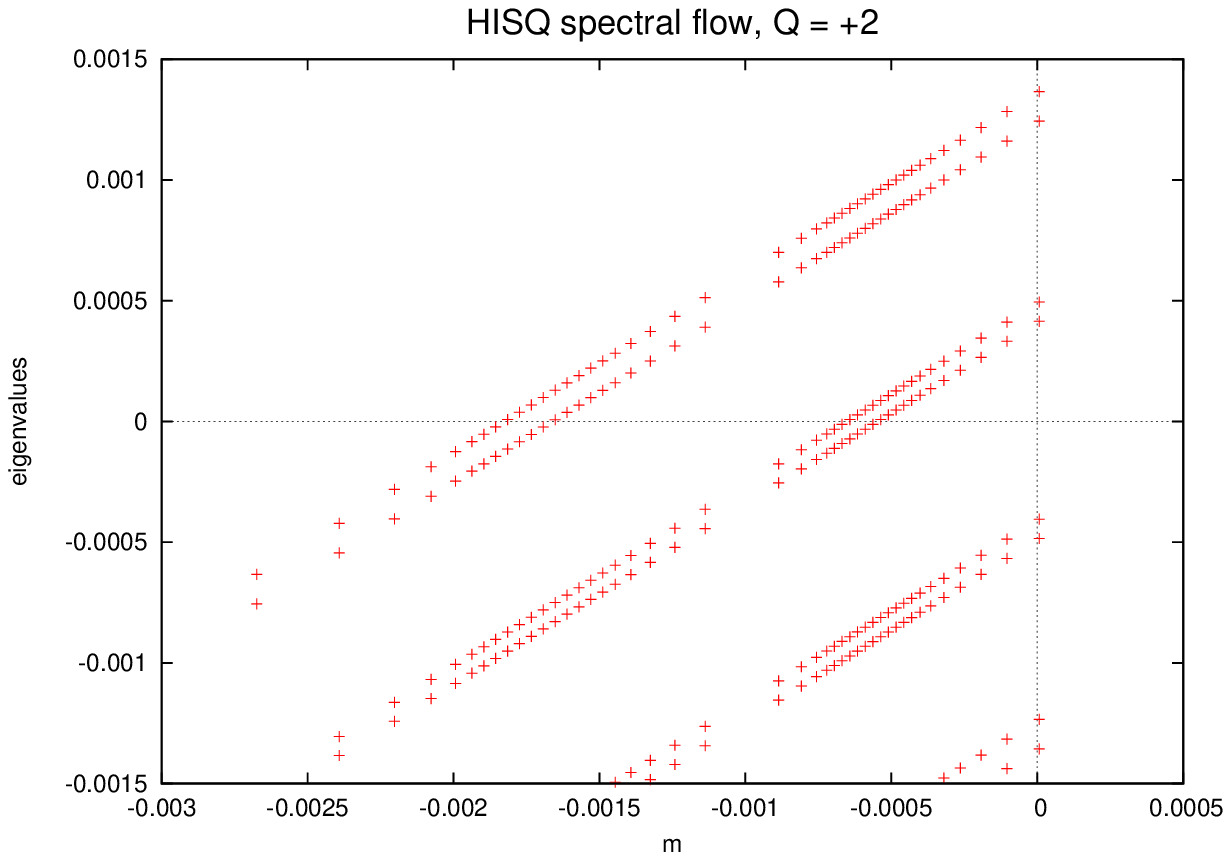} 
\end{array}
\]
\end{center}
\caption {Top left figure: taste-singlet chirality for the low-lying
  modes of the HISQ Dirac operator. Top right and bottom figures:
  spectral flow for the low-lying modes of the corresponding hermitian
  operator $H_{st}(m)$, for various ranges of $m$. This is for a gauge
  configuration with $Q = +2$.}
\label{fig3}
\end{figure}

We show in figures \ref{fig1}, \ref{fig2} and \ref{fig3} the results
obtained for three configurations corresponding (a posteriori), to
topological charge 0, -1, and 2. We can clearly see the agreement
between both definitions of the topological charge, with the expected
$4Q$ high-chirality modes, and $4Q$ crossings at low $m$.

\begin{figure}[h]
\begin{center}
\[
\includegraphics[scale = .9]{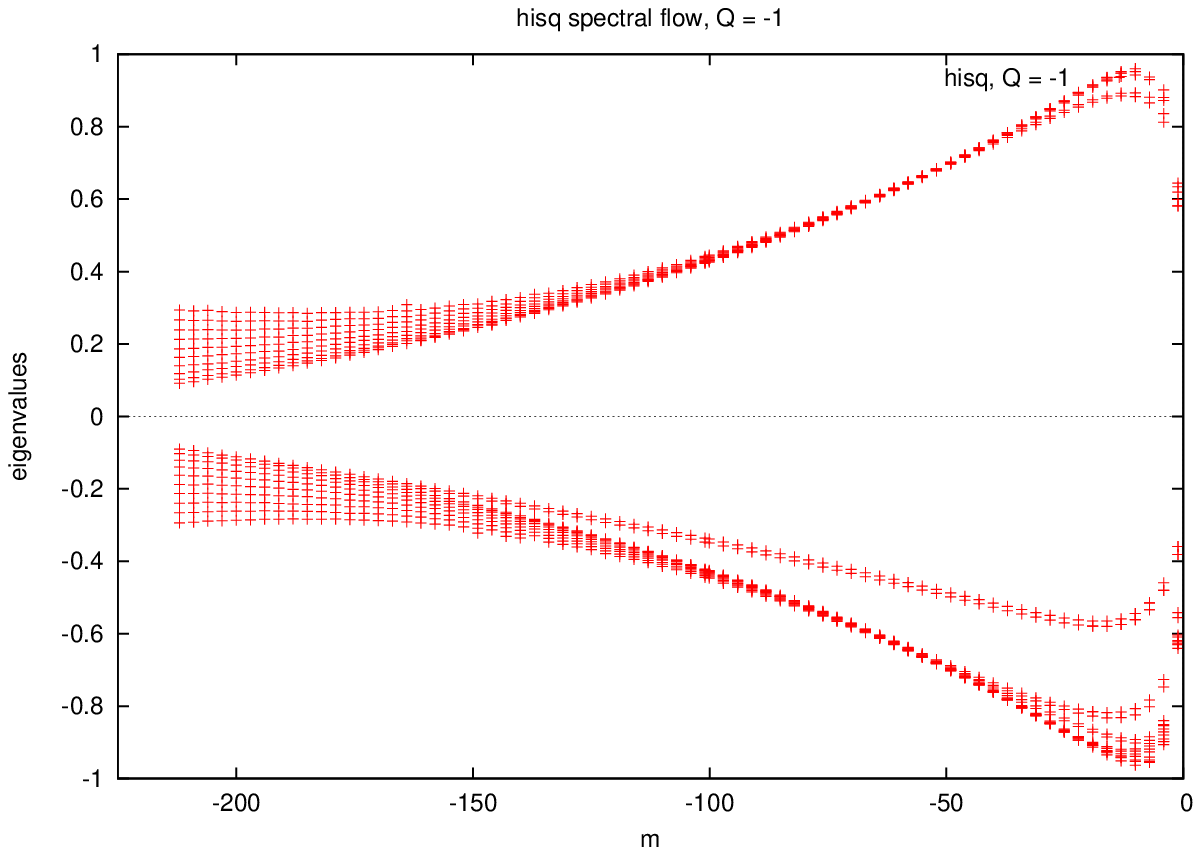} 
\]
\end{center}
\caption{Spectral flow corresponding to the HISQ Dirac operator on a
  large mass range, for a gauge configuration with $Q = -1$.}
\label{fig_long}
\end{figure}

In order for the topological charge to be well defined by the spectral
flow, it is necessary that the crossings at low values of the mass and
other possible crossings at larger values of the mass are well
separated. We show in figure \ref{fig_long} the spectral flow for the
same gauge configuration corresponding to $Q = -1$, but with a much
larger mass range. We see that there is no sign of any other crossing
until a very large value of $m$, of order ${\cal O}(200)$. We conclude
that, at least at this lattice spacing and for the HISQ Dirac
operator, there is a very good separation between low and high mass
crossings.

\begin{figure}[h]
\begin{center}
\[
\begin{array}{cc}
\includegraphics[scale = .65]{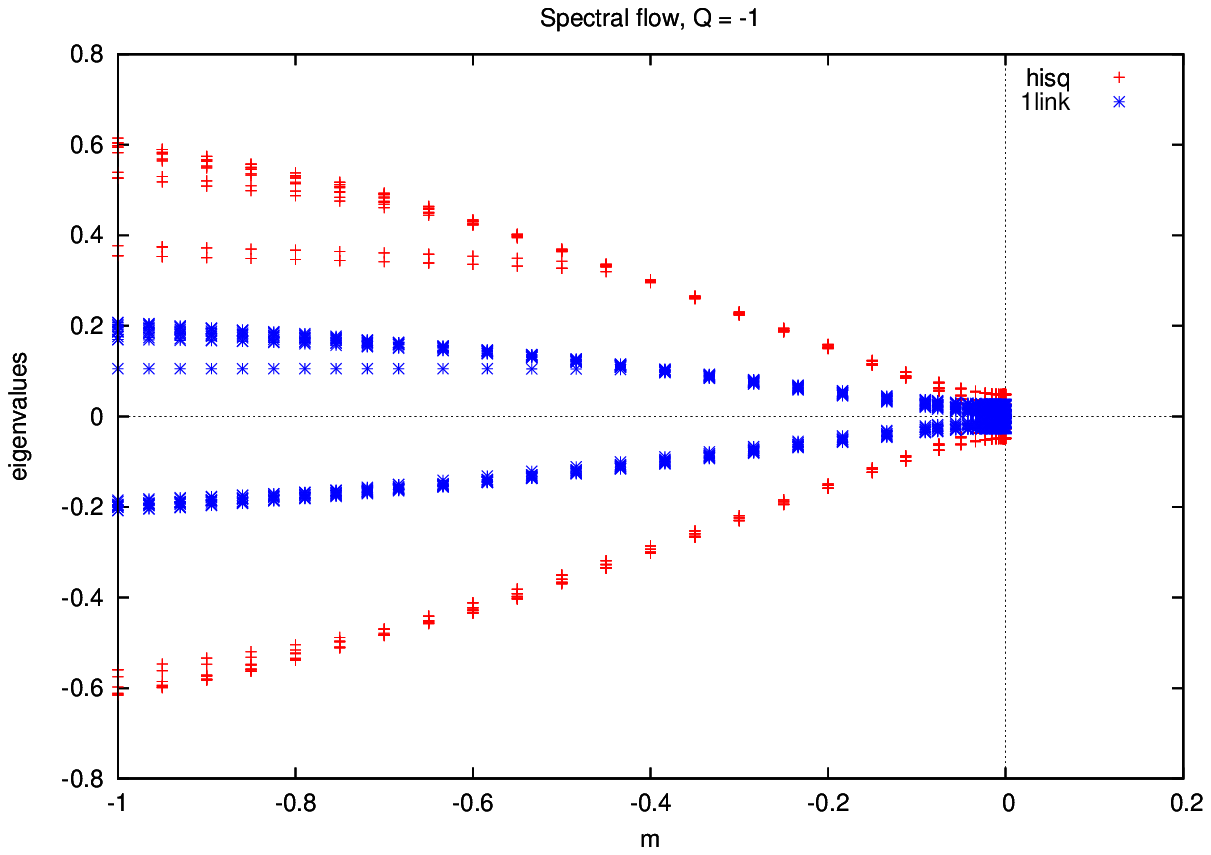} &
\includegraphics[scale = .6]{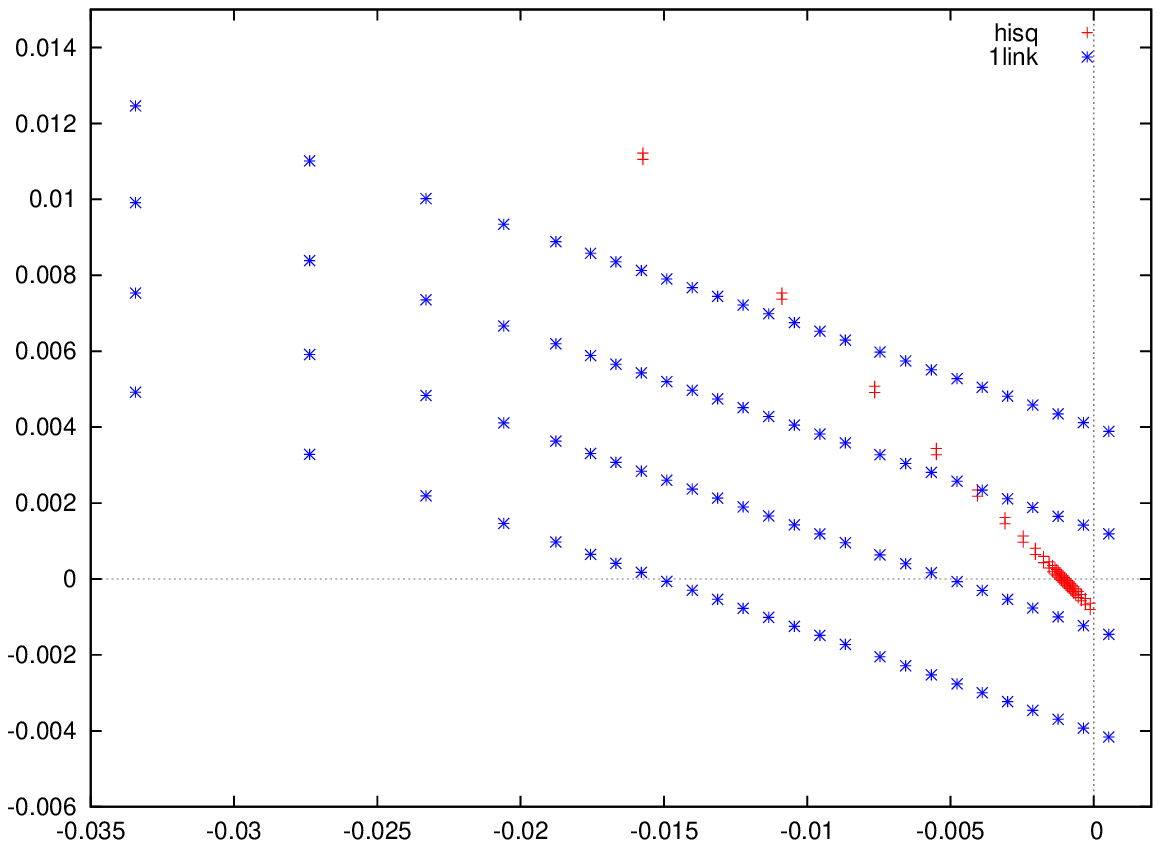} 
\end{array}
\]
\end{center}
\caption{The spectral flow corresponding to the HISQ and the 1-link
  Dirac operators, on the same gauge field configuration.}
\label{fig1link}
\end{figure}

In figure \ref{fig1link} we compare the spectral flow coming from the
HISQ and the 1-link Dirac operators, on the same gauge field
configuration, of topological charge -1. Both flows agree on the value
of the topological charge of the configuration, but the crossings
corresponding to HISQ take place at a much smaller value of $m$. This
is according to expectations, because in the continuum limit the only
possible crossing is at $m = 0$, and we expect the HISQ operator to be
much closer to the continuum than the 1-link operator. Another
manifestation of this is the fact that the four-fold degeneracy of the
continuum theory is also much more closely approximated by the HISQ
action, due to its much reduced taste-symmetry breaking.

\begin{figure}[h]
\begin{center}
\[
\includegraphics[scale = .6]{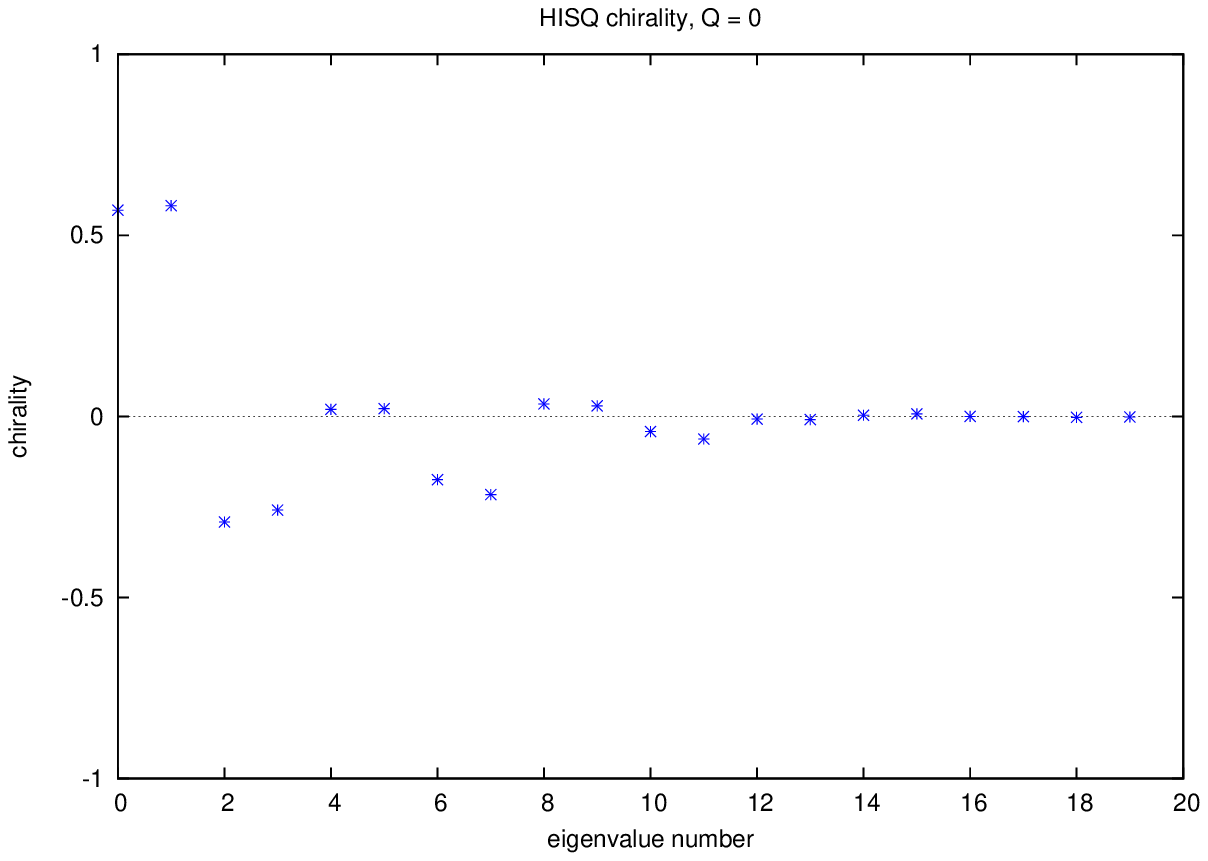}
\]
\[
\begin{array}{cc}
\includegraphics[scale = .6]{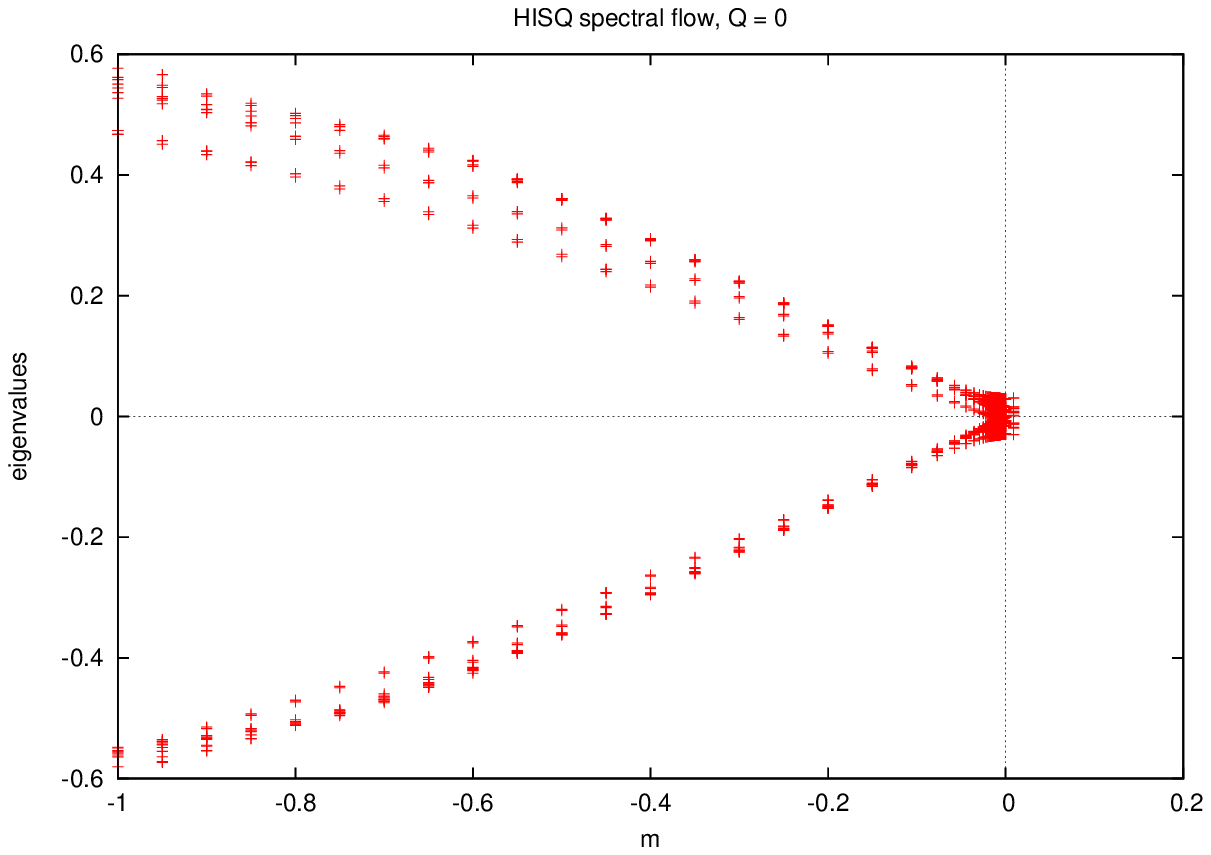} &
\includegraphics[scale = .6]{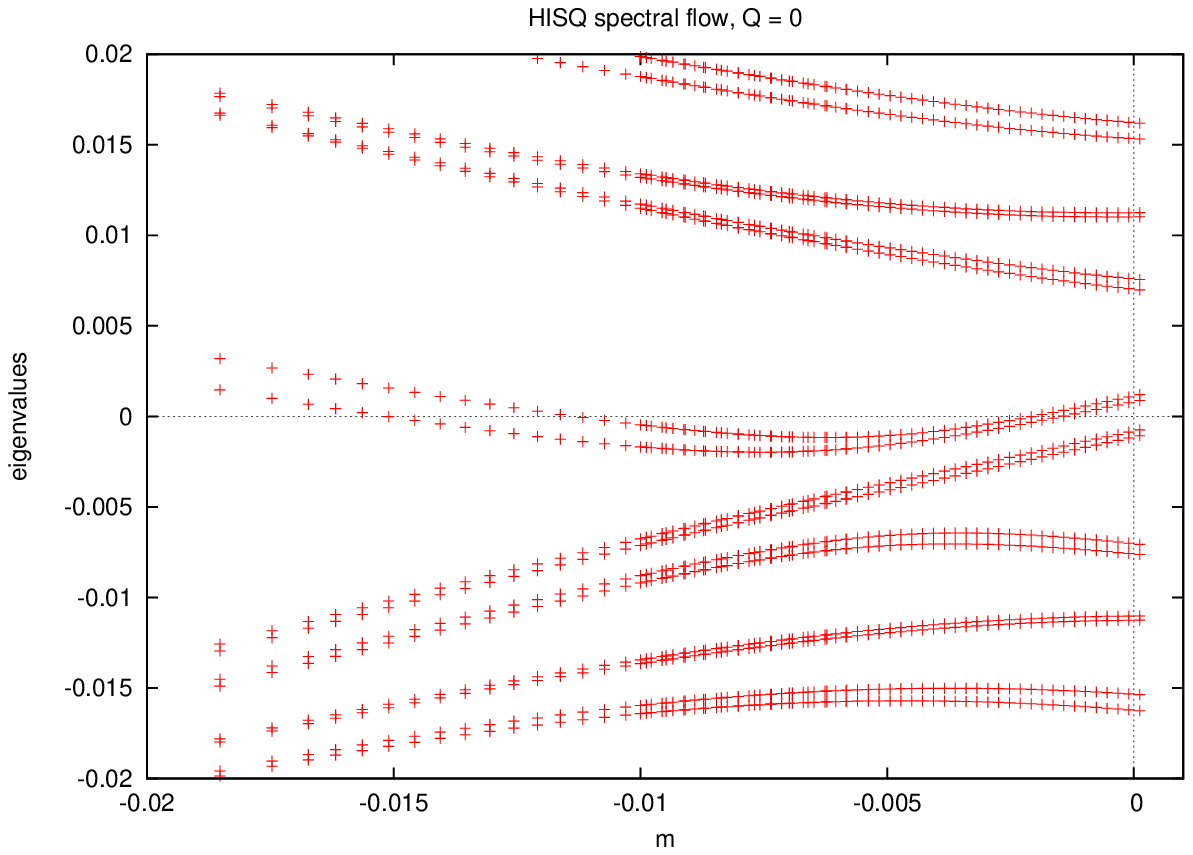} 
\end{array}
\]
\end{center}
\caption{Chiralities and HISQ spectral flow for a configuration with
  an ambiguous topological charge, as determined by the chiralities.}
\label{figpato}
\end{figure}

The definition of topological charge through the identification of the
high-chirality, low-lying modes of the Dirac operator works well in
practice, and any ambiguities are expected to vanish in the continuum
limit as $a^2$, where $a$ denotes the lattice spacing. Nevertheless,
at finite lattice spacing there are a few configurations for which the
classification in a topological sector is not clear-cut \cite{top1}.
In figure \ref{figpato} we show the chiralities and spectral flow for one
of those configurations. The high-chirality criterion would indeed be
ambiguous applied to this configuration. The spectral flow criterion
is always well-defined\footnote{At least as long as there is a
  clear-cut separation between low-modes and high-modes.}, and would
assign a topological charge 0 to this configuration. We can see,
however, that this is the result of having pairs of crossings with
opposite slopes, instead of not having any crossing (as is the case in
figure \ref{fig1} for the configuration with $Q = 0$).

\section{Conclusions and Outlook}

We have presented preliminary numerical evidence that Adams'
definition of the topological charge using the staggered Dirac
operator works as expected for realistic (quenched) $SU(3)$ gauge
fields. The crossings corresponding to low and high-lying modes are
well separated, and therefore the topological charge of a
configuration is unambiguously defined, even in cases which are
ambiguous using other definitions.

It would be interesting to compare the staggered Dirac spectral flow
with the usual Wilson Dirac spectral flow on the same gauge
configurations, as well as studying the dependence on the lattice
spacing \cite{futuro}.

Inspired by this definition of the spectral flow, one can define an
overlap operator starting with a staggered kernel, instead of the
usual Wilson one \cite{Adams2}, producing a chiral operator
representing two tastes of fermions. A similar construction can be
carried out to further reduce the degeneracy and produce a one-flavour
overlap operator \cite{Hoelbling}. The question now is whether this
construction is numerically advantageous as compared with the usual
overlap construction. Preliminary results are presented in
\cite{deForcrand}.

\section{Acknowledgments}

We thanks Alistair Hart for generating the configurations.  This work
was funded by an INFN-MICINN collaboration, MICINN (under grants
FPA2009-09638 and FPA2008-10732), DGIID-DGA (grant 2007-E24/2), and by
the EU under ITN-STRONGnet (PITN-GA-2009-238353). E. Follana is
supported on the MICINN Ram\'on y Cajal program. and A. Vaquero was
supported by MICINN through the FPU program.


\begin{thebibliography}{99}

\bibitem{Adams1}
  D.~H.~Adams,
  Phys.\ Rev.\ Lett.\  {\bf 104 } (2010)  141602.
  [arXiv:0912.2850 [hep-lat]].

\bibitem{futuro}
V.~Azcoiti, G.~Di~Carlo, E.~Follana, A.~Vaquero,
  in preparation.

\bibitem{Golterman}
  M.~F.~L.~Golterman,
  Nucl.\ Phys.\  {\bf B273 } (1986)  663.

\bibitem{hisq}
  E.~Follana Q.~Mason, C.T.H.~Davies, K.~Hornbostel, 
  G.P.~Lepage, J.~Shigemitsu, H.~Trottier, K.~Wong,
  Phys.\ Rev.\  {\bf D75 } (2007)  054502.
  [hep-lat/0610092].

\bibitem{top1} 
E.~Follana, A.~Hart, C.T.H.~Davies, 
Phys.\ Rev.\ Lett.\ {\bf 93 } (2004) 241601.
  [hep-lat/0406010].

\bibitem{top2} 

E.~Follana, A.~Hart, C.T.H.~Davies, Q.~Mason, 
Phys.\ Rev.\ {\bf D72 }
(2005) 054501.  [hep-lat/0507011].




  
\bibitem{Adams2}
  D.~H.~Adams,
   Phys.\ Lett.\  {\bf B699 } (2011)  394-397.
  [arXiv:1008.2833 [hep-lat]].

\bibitem{Hoelbling}
  C.~Hoelbling,
  Phys.\ Lett.\  {\bf B696 } (2011)  422-425.
  [arXiv:1009.5362 [hep-lat]].

\bibitem{deForcrand}
  P.~de Forcrand, A.~Kurkela, M.~Panero,
  PoS {\bf LATTICE2010 } (2010)  080.
  [arXiv:1102.1000 [hep-lat]].

\end{thebibliography}
\end{document}